\documentclass[jkps,preprint,twocolumn,fleqn,showpacs,showkeys]{revtex4}
\usepackage{graphicx}
\usepackage{amssymb}
\usepackage{amsmath}
\usepackage{bm}
\begin{document}
\setcounter{page}{0}
\title[]{Gamow's Calculation of the Neutron Star's Critical Mass Revised}
\author{Hendrik \surname{Ludwig}}
\email{hendrik.ludwig@icranet.org}
\author{Remo \surname{Ruffini}}
\email{ruffini@icra.it}
\affiliation{Dipartimento di Fisica and ICRA, Sapienza Universit\`a di Roma\\
P.le Aldo Moro 5, I-00185 Rome, Italy\\
ICRANet, P.zza della Repubblica 10, I-65122 Pescara, Italy\\
ICRANet, University of Nice-Sophia Antipolis, 28 Av. de
Valrose, \\
06103 Nice Cedex 2, France}

\date{\today}

\begin{abstract}
It has at times been indicated that Landau introduced neutron stars in his classic paper of 1932.
This is clearly impossible because the discovery of the neutron by Chadwick was submitted more than one month after Landau's work.
Therefore, and according to his calculations, what Landau really did was to study white dwarfs, and the critical mass he obtained clearly matched
the value derived by Stoner and later by Chandrasekhar.
The birth of the concept of a neutron star is still today unclear. Clearly, in 1934, the work of Baade and Zwicky pointed to neutron stars as originating from supernovae.
Oppenheimer in 1939 is also well known to have introduced general relativity (GR) in the study of neutron stars.
The aim of this note is to point out that the crucial idea for treating the neutron star has been advanced in Newtonian theory by Gamow. However, this pioneering work was
plagued by mistakes. The critical mass he should have obtained was $6.9\,M_\odot$, not
the one he declared, namely, $1.5\ M_\odot$. Probably, he was taken to this result by the work of Landau on white dwarfs.
We revise Gamow's calculation of the critical mass regarding
calculational and conceptual aspects and discuss whether it is justified to consider it the first neutron-star critical mass.
We compare Gamow's approach to other early and modern approaches to the problem.
\end{abstract}

\pacs{04.25.Dm, 04.40.Dg, 97.60.Jd}

\keywords{Critical mass, History of astrophysics}

\maketitle

\section{INTRODUCTION}

In Ref. \cite{gamow} Gamow attempted to calculate the critical mass of a neutron star like
object, approximating it with a homogenous matter distribution, as had previously been done by Stoner \cite{stoner1} in the context of white dwarfs.
Besides conceptual flaws, the derivation contains some calculational mistakes. The purpose of this work is to correct the
latter and discuss the former. Gamow describes a state of matter at densities
$\rho \approx 10^{12}\mathrm{g/cm^3}$, where electrons have been absorbed by the $\beta$-process and
nuclear exchange forces come into play, which he ignores for the remainder of the calculation.
He refers to this state as the {\it nuclear state} and calls stellar cores in this state
{\it stellar nuclei}. The idea of Gamow's calculation is to compare the dependences of the
Newtonian gravitational pressure
\begin{align}
P_g=-\frac{\partial U_g}{\partial V} 
\end{align}
and the non- and the ultra-relativistic degeneracy pressures
\begin{align}
P_F=-\frac{\partial E_F}{\partial V},\ \ 
P_{F,R}=-\frac{\partial E_{F,R}}{\partial V}
\end{align}
on the mass density $\rho$. They are obtained for a homogeneous mass distribution, which
appears to be a reasonable approximation for an order of magnitude estimate. Note that
in this notation $P_F$ stands for the pressure of the Fermi gas and is not to be confused with
the Fermi momentum. The total energies,
including the rest-mass and the internal energies, but excluding the gravitational self-energy, are denoted
$E_F$ and $E_{F,R}$, respectively, again not to be confused with the Fermi energy. In his
calculation, Gamow obtains
\begin{align}
 P_g\propto\rho^{4/3},\ P_F\propto\rho^{5/3},\ P_{F,R}\propto\rho^{4/3},
\end{align}
and he concludes that for a non-relativistic Fermi gas, the degeneracy pressure will always
balance the gravitational pressure as soon as a sufficiently high mass density is reached.

In the derivation of $P_F$, he assumes proportionality of mass density and number density, which is
correct for a non-relativistic gas of Fermions. He also concludes that the fate of an
ultra-relativistic mass distribution depends on the ratio $P_G/P_{F,R}$ at some initial time, as
it will stay constant during a potential collapse. The flaw in his argument is the
assumption of propotionality between the mass density and the number density in the derivation of
$P_{F,R}$, which, for a free neutron gas, is not correct. In Section \ref{sec:concFlaw}, we will
argue that this assumption holds approximately true for a white dwarf with Fermi energy between the electron and deuterium mass regions, $m_e c^2\ll E_\mathrm{Ferm} \ll 2 m_N c^2$.
In the next section, we will turn to the calculational and conceptual flaws present in Ref. \cite{gamow} and correct the
numerical values obtained.

\begin{table}
 \begin{center}
\begin{tabular}{ccc}
 \multicolumn{3}{c}{Homogenous models}\\
 \hline
 &$\rho_\mathrm{non}\propto n$ & $\rho_\mathrm{ult}\propto n^{4/3}$\\ \hline\hline
$p_\mathrm{non}\propto n^{5/3}$\ \ \  &\ \ no critical mass\ \ & not considered\\
$p_\mathrm{ult}\propto n^{4/3}$\ \ \  & $6.9\ M_\odot$ & no critical mass\\ \hline
\end{tabular}
\end{center}
\label{tab:homogen}
\caption{Critical masses in homogenous Newtonian models for all combinations of non- and ultra-relativistic approximations
of $p$ and $\rho$.}
\end{table}

\begin{figure*}
\includegraphics[width=14.0cm]{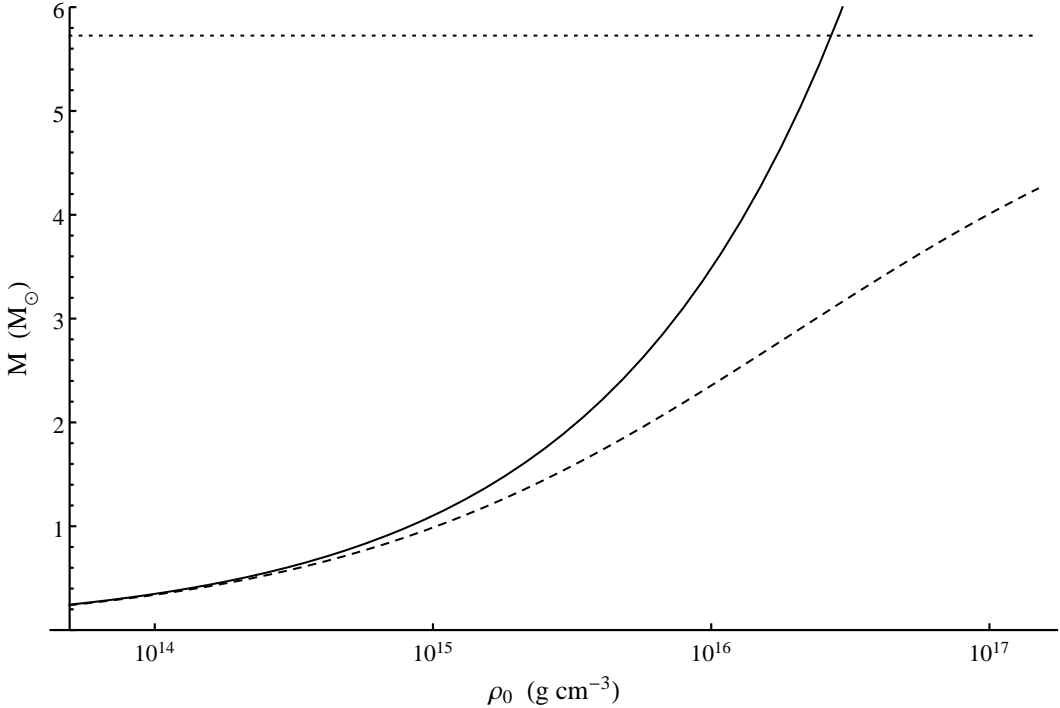}
 \label{fig:hydroNaive}
 \caption{Relation between the total mass $M$ and the central density $\rho_0$ for a non-relativistic mass density $\rho$ and non-relativistic (solid),
 fully-relativistic (dashed) and ultra-relativistic (dotted) pressure $p$. The second model interpolates between the non- and the ultra-relativistic ones.}
\end{figure*}

\section{REVISING THE CALCULATION}\label{sec:concFlaw}
Counting the states of a Fermi gas, Gamow introduces an erroneous factor of $\pi$.
He obtains the degeneracy pressure of the Fermi gas by counting states in a cube of zero
potential and side length $l$, with infinite potential walls at the boundary, for both the non-
and ultra-relativistic cases. He claims that, for large
$n$, the number of states sharing the same value
\begin{align}
 n=\sqrt{{n_x}^2+{n_y}^2+{n_z}^2}
\end{align}
is approximately given by $n^2$. This is not correct and should be replaced by $\pi\, n^2$.
When the non-relativistic
degeneracy pressure is derived, this error is canceled by another error and leads him to the correct (see Ref. \cite{landau}) expression
\begin{align}
 P_F&=\frac{3^{2/3}\pi^{4/3}\hbar^2}{5 m^{8/3}}\rho^{5/3}.
\end{align}
In the case of the ultra-relativistic degeneracy pressure, the initial error persists to the end of the calculation.
The (formally) correct result (see Ref. \cite{landau}) reads
\begin{align}\label{eqn:ultraRelPress}
 P_{F,R} &=\frac{3^{1/3}\pi^{2/3}\hbar c}{4}\left(\frac{N}{l^3}\right)^{4/3}\nonumber\\
 &=\frac{3^{1/3}\pi^{2/3}\hbar c}{4 m^{4/3}}\rho^{4/3}
\end{align}
and differs from Gamow's result by a factor of $3\,\pi^{1/3}$. This is, however, based on the assumption
\begin{align}\label{eqn:rhoProp_n}
 \rho&=m N/l^3 \propto n,
\end{align}
i.e., the assumption that the energy density is provided only by the rest mass of the particles and is, thus, proportional to the number density $n$.
This assumption is not valid for a free gas of neutrons,
but holds only if the pressure is provided by a light, ultra-relativistic particle like, e.g., an electron, while the energy density is provided by a
heavy, non-relativistic particle like, e.g., a proton or a deuteron, a situation that is found in white dwarfs.

Equation (\ref{eqn:ultraRelPress}) is to be compared to the expression that Gamow provides for the
gravitational pressure and which reads
\begin{align}\label{eqn:gravPress}
 P_g&=\frac{3}{20\pi}k\frac{M^2}{R^4}\nonumber\\
 &=\frac{1}{5}\left(\frac{4\pi}{3}\right)^{1/3}k M^{2/3}\rho^{4/3},
\end{align}
where $k$ denotes the gravitational constant. This expression is correct within the Newtonian
framework, and it is not per se inconsistent with the possibility that the energy density $\rho$
depends on the volume of the sphere even for a fixed particle number. One has to note, however,
that this expression does not take into account the effect of the gravitational self-energy itself on
the total mass. This would lead to a nonlinear problem and was in fact the main motivation for
Einstein \cite{einstein} to develop his famous General Theory of Relativity, which overcomes the
inconsistencies between Newton's gravity and the Special Theory of Relativity. Ignoring these
self-energy constributions for now and demanding that the gravitational pressure be higher than the degeneracy
pressure leaves us with the inequality
\begin{align}
 \frac{1}{5}\left(\frac{4\pi}{3}\right)^{1/3}k M^{2/3}\rho^{4/3}\nonumber\\
 >\frac{3^{1/3}\pi^{2/3}\hbar c}{4 m^{4/3}}\rho^{4/3},
\end{align}
where the right-hand side is already based on the assumption in Eqn. (\ref{eqn:rhoProp_n}), which is only
valid for a white dwarf in a region where pressure is provided by ultra-relativistic electrons
and the mass density is provided by non-relativistic baryons.

We will have to embrace Eqn. (\ref{eqn:rhoProp_n})
another time to deduce that the total mass $M$ is a constant during collapse and does not depend on
the mass density $\rho$ as long as the total number of particles is conserved. This allows us to
remove $\rho$ from the inequality and provides us with the expression
\begin{align}\label{eqn:critMass}
 M_\mathrm{crit}=\sqrt{5\pi}\frac{15 {m_p}^3}{16 m^2}
\end{align}
for the critical mass, where $m_p=\sqrt{\hbar c/k}$ is the Planck mass.
The assumption in Eqn. (\ref{eqn:rhoProp_n}) is, as we have stated before, not valid when considering a
degenerate Fermi gas consisting of a single constituent as is the case in the free-neutron-gas model
for a neutron star. This is because the gas can either be in the ultra-relativistic region, necessary
for Eqn. (\ref{eqn:ultraRelPress}) to hold or in the non-relativistic region, necessary for Eqn.
(\ref{eqn:rhoProp_n}) to hold, but not in both at the same time.

The ultra-relativistic approximation is
based on the fact that the rest mass can be neglected with respect to the kinetic energy, so one is
underestimating the mass density significantly when counting only rest masses. If we consider, however,
a white dwarf consisting of electrons and baryons with certain mass number $A$ and atomic number $Z$
and assume local charge neutrality, eliminating all electric fields, we can construct a situation where the
number densities of electrons and baryons are related by $n_B=Z\,n_e$, and due to the big difference in
rest masses between electrons and baryons, this system can be in a state where electrons, providing the pressure,
are at ultra-relativistic densities while baryons, providing the mass density, are at non-relativistic
densities. This would allow us to predict a partial collapse, but we have to keep in
mind that at later stages of this process, baryons will become ultra-relativistic, $\beta$-processes will
begin to play a role, and, as soon as nuclear densities are reached, the strong force will come into play.

We can calculate numerical values for the critical mass in Eqn. (\ref{eqn:critMass}), either by substituting the
proton mass $m\rightarrow m_\mathrm{proton}=938.3\,\mathrm{MeV\,c^{-2}}$, which leads to
\begin{align}\label{neutronCritMass}
 M_\mathrm{crit,p}\approx 6.89\,M_\odot,
\end{align}
but represents a white dwarf consisting of electrons and protons only, or by substituting the deuteron
mass $m\rightarrow m_\mathrm{deuteron}=1.876\,\mathrm{GeV\,c^{-2}}$, which leads to
\begin{align}
 M_\mathrm{crit,D}\approx 1.72\,M_\odot,
\end{align}
and roughly represents a white dwarf consisting of electrons and a type of baryons fulfilling the
approximate relation $Z\approx A/2$.

In Ref. \cite{gamow} Gamow obtains a value of $1.5\,M_\odot$, which deviates
considerably from (\ref{neutronCritMass}), the value he attempted to compute in the framework of his method.
The discrepancy cannot be accounted for by the numerical error mentioned above and must be based on
further calculational mistakes or on a carry over of the value obtained, e.g., by Landau \cite{landau1}, one month
before the discovery of the neutron \cite{chadwick}.

\begin{table}
 \begin{center}

\begin{tabular}{ccc}
 \multicolumn{3}{c}{Hydrostatic models}\\
 \hline
 &$\rho_\mathrm{non}\propto n$ & $\rho_\mathrm{ult}\propto n^{4/3}$\\ \hline
$p_\mathrm{non}\propto n^{5/3}$\ \ \  &\ \ no critical mass\ \ & not considered\\
$p_\mathrm{ult}\propto n^{4/3}$\ \ \  & $5.7\ M_\odot$ & infinite radius\\ \hline
\end{tabular}
\end{center}
\label{tab:hydro}
\caption{Critical masses in hydrostatic Newtonian models for all combinations of non- and ultra-relativistic approximations
of $p$ and $\rho$.}
\end{table}

\section{DISCUSSION}\label{sec:disc}

\begin{figure*}
\includegraphics[width=14.0cm]{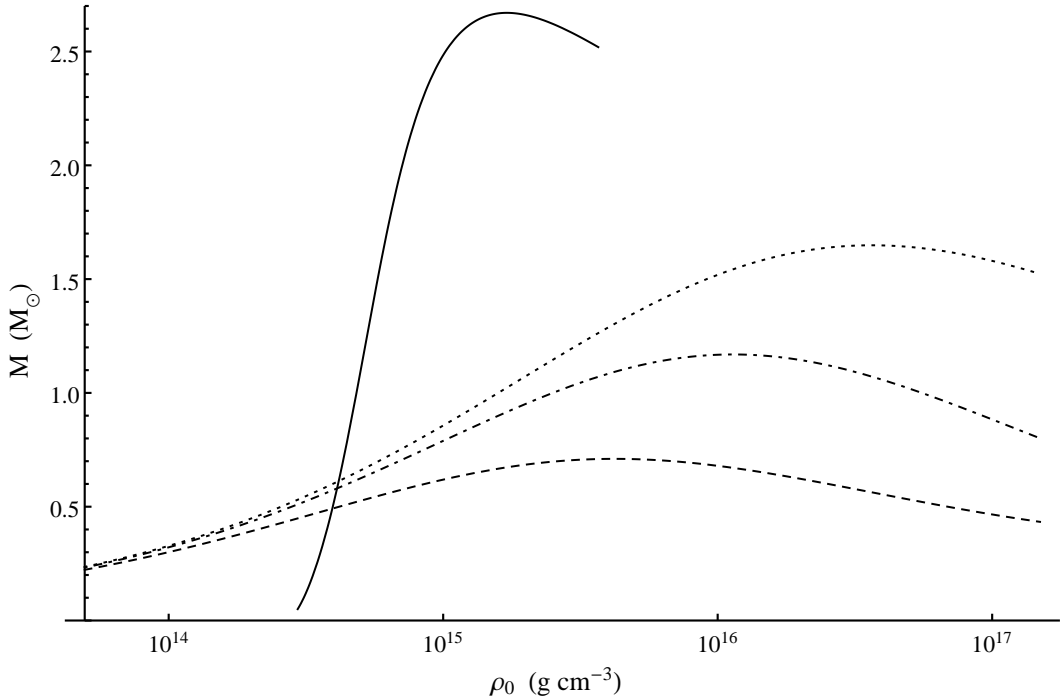}
 \caption{Relation of the total mass $M$ to the central density $\rho_0$ for a fully-relativistic mass density $\rho$ and fully-relativistic pressure $p$. We compare
 the strictly Newtonian model (solid), the Newtonian model with gravitational self-energy corrections (dashed), and the general relativistic model (dotted)
 to the state-of-the-art model treated in Ref. \cite{rueda}.
 Coincidentially, the critical mass of the first model lies near the white-dwarf's critical mass.}
 \label{fig:hydroFull}
\end{figure*}

In Fig. \ref{fig:hydroNaive} we present a comparison of three simplified neutron-star models, plotting the
respective relations between mass and central density. They all use a non-relativistic ($\rho\propto n$) energy density and
are based on a balance between the Newtonian gravitational force and the hydrostatic pressure force.
The solid line represents a non-relativistic degeneracy pressure and exhibits no citical mass, as the total mass
monotonically depends on the central pressure. The dotted line is based on an ultra-relativistic degeneracy pressure,
and this model has the perculiar property that the whole family of solutions has the same total mass of $5.7\,M_\odot$, regardless of the central density.
The dashed line represents a full relativistic expression for the degeneracy pressure,
\begin{align}\label{fullPress}
 p=&\frac{c^5 m_N^4}{8\pi^2\hbar^3}\Big[\sqrt{1+\eta^2}(-\eta+\frac{2}{3} \eta^3)\nonumber\\
 &\ \ \ \ \ \ \ \ \ \ \ \ \ \ \ \ \ \ \ \ \ \ \ \ \ +\mathrm{arsinh}\,\eta\Big],\nonumber\\
 \eta=&(3\pi^2 n)^{1/3}\frac{\hbar}{c\,m_N},
\end{align}
and, accordingly, interpolates
between the former two, asymptotically approaching the critical mass of $5.7\,M_\odot$, but lacking an unstable branch provided by the models
presented in Fig. \ref{fig:hydroFull}.

In Tables \ref{tab:homogen} and \ref{tab:hydro}, we compare the results of the homogenous method
used by Gamow and the hydrostatic model used in Fig.\ref{fig:hydroNaive}. The two critical masses of $6.9\,M_\odot$ and $5.7\,M_\odot$ are
close, and both are caused by the pressure loss during transition from a non- to an ultra-relativistic pressure. The relatively small difference might be considered surprising though,
because the homogenous model has to reach an ultra-relativistic pressure in the entire volume while the hydrostatic model has to do so
only at the center.
In Fig. \ref{fig:hydroFull}, we present a comparison of the three simplified neutron-star models that use the fully-relativistic expressions for
the energy density,
\begin{align}
 \rho=&\frac{c^5 m_N^4}{8\pi^2\hbar^3}\Big[\sqrt{1+\eta^2}(\eta+2\eta^3)\\
 &\ \ \ \ \ \ \ \ \ \ \ \ \ \ \ \ \ \ \ \ \ \ \ \ \ -\mathrm{arsinh}\,\eta\Big],\nonumber
\end{align}
and pressure, Eqn. (\ref{fullPress}), of a degenerate Fermi gas, as was first done by Oppenheimer and Volkoff \cite{oppenVol} in the context of neutron stars, as well as
the state-of-the-art model in Ref. \cite{rueda}.
The dotted line is again based on the strictly Newtonian hydrostatic model while the dash-dotted line incorporates (negative) gravitational self-energy,
and the dashed line was obtained in the full General Relativistic treatment of a degenerate neutron gas, i.e., the Tolman-Oppenheimer-Volkoff equation \cite{tolman,oppenVol}.
These three lines closely resembles the results of Rees et al. in \cite{ruffini} and differ only in the respect
that in their work the Harrison-Wheeler equation of state is used, which assumes a mix of
fully relativistic Fermi gases of electrons, protons, and neutrons in $\beta$-equilibrium, thus fulfilling
the relation $E^F_e+E^F_p=E^F_n$ between their respective Fermi momenta.

From the three simple models in Fig. \ref{fig:hydroFull}, we obtain critical masses of $1.6\,M_\odot$, $1.2\,M_\odot$, and $0.7\,M_\odot$, respectively,
which significantly differ from Gamow's corrected result, Eqn. (\ref{neutronCritMass}).
Also, in this case, the instability has to be attributed to the pressure loss when entering the ultra-relativistic region, but the very significant
difference in the critical masses obtained in Fig. \ref{fig:hydroNaive} shows that this region is reached at considerably smaller masses when
the internal energy of the neutron gas is take into account. The effect of the Fermion rest mass on the degeneracy pressure, on the other hand, is insignificant,
because it does not play a role in ultra-relativistic regions. Provided the number density is high enough, electrons and neutrons provide the same degeneracy pressures.\\
The values for the critical mass obtained from the fully-relativistic equation of state can still not be considered
to be applicable to astrophysics, because they do not take into account a crust, a rotation, or a strong nuclear force, which tend
to increase the critical mass to about $M_\mathrm{crit}\approx 2.7\,M_\odot$ \cite{rueda,belvedere}; see the solid line in Fig. \ref{fig:hydroFull}.
The strong nuclear force or residual strong force, which is also discussed by Gamow in his work, is mediated by mesons and changes the
equation of state considerably because it is attractive in certain regions, but repulsive in others, the transition occuring at an average
separation of about $0.7\,\mathrm{fm}$ or a number density of about $4.8\,\mathrm{mol/cm^3}$ \cite{wiringa}.
However, most importantly, there has been a breakthrough in recent years \cite{rotondo,rueda1} in understanding that the assumption of local charge neutrality, i.e., that the proton
and the electron densities coincide at each point, leads to inconsistencies regarding the constancy of the Klein potentials. This assumption has to be dropped in favor of global
charge neutrality, which admits electric fields within the matter distribution, but demands that they vanish in the exterior \cite{rueda,belvedere}.
For a review on other modern neutron star models, see Ref. \cite{lattimer}, where modifications like two- and three-body interactions, spin-orbit coupling,
nuclear stability and compressibility, as well as surface and symmetry energy, are introduced.

\section{CONCLUSIONS}
The fact that numerical errors are present in Gamow's calculation alone would not suffice to consider
it a failed attempt to calculate the neutron-star's critical mass, as it is an order of magnitude estimate,
and does not take into account a radius-dependent density profile. However, because his approach is not applicable to
the case of a free neutron gas, but rather to white dwarfs with ultra-relativistic electron pressures, and because he obtains
the value of $1.5\,M_\odot$ by coincidence, while the corrected result reads $6.9\,M_\odot$, his result cannot
be considered the first estimate of a neutron-star's critical mass. It should be considered a reproduction of earlier works
\cite{stoner1,anderson, stoner, chandrasekhar,chandrasekhar1,landau1} of Anderson, Stoner, Chandrasekhar, and Landau in the context of white dwarfs.
The development of the first consistent neutron-star model should be attributed to Tolman, Oppenheimer, and Volkoff not only because
it involves general relativitiy, but especially because it takes internal-energy contributions into account.
Nevertheless, it was Gamow who initiated the transition of the neutron star from a theorized object \cite{zwicky} to an object
with predictable properties that could be tested by observation.

\begin{acknowledgments}
The authors thank Prof. Jorge Rueda and Prof. She-Sheng Xue for helpful discussions and
gratefully acknowledge the hospitality at the APCTP where part of this work was done.\\
H. Ludwig is supported by the Erasmus Mundus Joint Doctorate Program under Grant
Number 2012-1710 from the EACEA of the European Commission.
\end{acknowledgments}

\bibliographystyle{unsrt}

\end{document}